\documentclass[a4paper]{article}
\usepackage[latin1]{inputenc}
\usepackage{amscd}
\usepackage[english]{babel}
\usepackage{amsmath,amssymb,amsfonts}
\usepackage{syntonly}

\begin{document}

\title{Blocking Sets in the complement of hyperplane arrangements in projective space}
\author {
S. Settepanella \thanks {Dipartimento di Scienze della Comunicazione,Colle Parco, 64100 Teramo, Italy}}
\date {January 2008}
\maketitle

\newcommand{\Q}{\mathbb Q}
\newcommand{\Z}{\mathbb Z}
\newcommand{\N}{\mathbb N}
\newcommand{\R}{\mathbb R}
\newcommand{\C}{\mathbb C}
\newtheorem{df}{Definition}
\newtheorem{theom}{Theorem}
\newtheorem{thm}{Theorem}
\newtheorem{prop}{Proposition}
\newtheorem{cor}{Corollary}
\newtheorem{lemma}{Lemma}
\newtheorem{rem}{Remark}
\newtheorem{example}{Example}


\newcommand{\PS}{\mathbf{P}=(\mathbf{S},\mathcal{P})}
\newcommand{\Pb}{\mathbf P}
\newcommand{\Ab}{\mathbf A}
\newcommand{\A}{\mathbf A}
\newcommand{\Sb}{\mathbf S}
\newcommand{\Pbp}{\mathbf P^{\prime}}
\newcommand{\Sp}{\mathcal S^{\prime}}
\newcommand{\Pcp}{\mathcal P^{\prime}}
\newcommand{\Pc}{\mathcal P}
\newcommand{\Wb}{\mathbf W}
\newcommand{\Kb}{\mathbb K}
\newcommand{\Gb}{\mathbf G}
\newcommand{\Sc}{\mathcal S}


\newcommand{\Kc}{\mathcal K}
\newcommand{\Kbf}{\mathbf K}
\newcommand{\Wc}{\mathcal W}
\newcommand{\Lc}{\mathcal L}
\newcommand{\Bc}{\mathcal B}


\newcommand{\Cal}{\mathcal}

\begin{abstract}
It is well know that the theory of minimal blocking sets is studied by 
several author. Another theory which is also studied by a large number of 
researchers is the theory of hyperplane arrangements.
We can remark that the affine space $AG(n,q)$ is the 
complement of the line at infinity in $PG(n,q)$. Then $AG(n,q)$ can be regarded
as the complement of an hyperplane arrangement in $PG(n,q)$!
Therefore the study of blocking sets in the affine space $AG(n,q)$ is 
simply the study of blocking sets in the complement of a finite 
arrangement in $PG(n,q)$.
In this paper the author generalizes this remark starting to study the problem 
of existence of blocking sets in the complement of a given hyperplane
arrangement in $PG(n,q)$. As an example she solves the problem for the case of 
braid arrangement.
Moreover she poses significant questions on this new and interesting 
problem.
\end{abstract}

\section{Introduction}

Throughout this paper, $PG(n,q)$ and $AG(n,q)$ will respectively denote 
the n-dimensional projective and affine space over the finite field 
$GF(q)$.

A \textit{t-blocking set B} in $PG(n,q)$ (or $AG(n,q)$) is a set $B$ of
points such that any $(n-t)$-dimensional subspace intersects $B$. A 
1-blocking set is simply called a \textit{blocking set}. A $t$-blocking 
set is called \textit{minimal}, if none of its proper subsets is also a 
$t$-blocking set. 

The smallest $t$-blocking sets have been characterized by Bose and Burton
\cite{BoseBurton}. They proved that the smallest $t$-blocking sets in 
$PG(n,q)$ are subspace of dimension $t$. An old result of Bruen 
\cite{Bruen1} states that the second smallest minimal blocking set in 
the plane $PG(2,\sqrt{q})$, $q$ a square, is (the point set) of a Baer 
subplane $PG(2,\sqrt{q})$.

There are several survey papers about blocking sets (see
Sz\~onyi, G\'acs, Weiner \cite{SzGaWe} and chapter 13 of the second 
edition of Hirschfeld's book \cite{Hirsc}). Most of the surveys 
concentrate on small
minimal blocking sets. But there are also several results about large 
minimal blocking sets. The first such result is due to Bruen and Thas 
\cite{BruenThas}. More results on the spectrum of minimal blocking sets 
in planes of small order can be found in Cossidente, G\'acs et alt 
\cite{CosSzGaSic} Innamorati \cite{Innam},Inamorati and Maturo \cite{Innamat}. 

Another interesting problem is the existence of minimal Blocking sets in
$PG(n,q)$ and $AG(n,q)$ for fixed $n$ and $q$ (see Mazzocca-Tallini 
\cite{173} and Beutelspacher-Eugeni \cite{82}).

It is well know that the theory of minimal blocking sets is studied by 
several author.
There is another theory which is also studied by a large number of 
researchers: the theory of hyperplane arrangements (see the Orlik-Terao's book
\cite{OS} for a survey on the theory).

An hyperplane arrangement is a set of hyperplane in a given space.

The author remarks that the affine space $AG(n,q)$ is the complement of
the line at infinity in $PG(n,q)$. Then $AG(n,q)$ can be regarded as the
complement of an hyperplane arrangement in $PG(n,q)$!

Therefore the study of blocking sets in the affine space $AG(n,q)$ is 
simply the study of blocking sets in the complement of a finite 
arrangement in $PG(n,q)$.

The author generalized this remark studying the problem of
the existence of blocking sets in the complement of a given hyperplane
arrangement in $PG(n,q)$.

In this paper author introduces the first results on this new theory,
she generalizes some results of the old theory to the new 
one. 

Moreover she solves the problem of existence of blocking sets in
the complement of a well-known and studied arrangement: 
\textit{the braid arrangement}, the set of reflection hyperplanes 
generated by the symmetric group.

She concludes the paper posing several questions:
\begin{enumerate}
\item on the existence of 
blocking sets on the complement of hyperplane arrangements;
\item on the characterization of hyperplane arrangements in $PG(n,q)$ 
which give rise to interesting new problem on blocking sets;
\item on the link between the two theories.
\end{enumerate}

Clearly also the problem of small and large blocking sets can be 
studied again in the complement of a given hyperplane arrangement.
The author is still working to this problem. 

Plainly in this paper author writes blocking sets 
instead of minimal blocking sets.

\section{General results on Blocking sets}

Let $PG(n,q)$ ($AG(n,q)$) the n-dimensional projective (affine) space 
over a finite field $GF(q)$. We have the following:

\begin{prop}\label{sottospazi} Let C be a blocking set in $PG(n,q)$ 
($AG(n,q)$) with $n > 2$ and $S_d$ a subspace of dimension $d>1$, then
the intersection $C \cap S_d$ is a blocking set in $S_d$.
\end{prop}

\begin{prop}\label{sovraspazi} Let $S_{n-1}$ an hyperplane in $PG(n,q)$ 
($n>2$). If $C_1$ is a blocking set in 
$AG(n,q)=PG(n,q) \setminus S_{r-1}$ and 
$C_2$ is a blocking set in $S_{r-1}$, then $C_1 \cup C_2$ is a blocking
set in $PG(n,q)$.
\end{prop}

\begin{cor}\label{somme}
If $PG(n,q)$ and $AG(n+1,q)$ contain blocking sets then also $\Pb \Gb (r+1,q)$ 
does.
\end{cor}

The proof of the above statements is trivial.

Mazzocca and Tallini in \cite{173} proved the following:

\begin{thm}\label{funzione importante}
There is a function $b_{p}(t,q)$
($b_a(t,q)$), depending on t and q, such that $PG(n,q)$ ($AG(n,q)$) 
contains t-blocking sets if and only if $r \leq b_p(t,q)$ 
($r \leq b_a(t,q)$).
\end{thm}

Another interesting general result on existence of blocking set is the
following:

\begin{thm}(A. Beutelspacher-F. Eugeni \cite{82}). 
Let $PG(n,q)$ and $AG(n,q)$ be the projective and affine n-dimensional 
space over a finite field $GF(q)$. If $q \geq 2^n$ then   
$AG(n,q)$ and $PG(n,q)$ contains t-blocking sets.
\end{thm}

\section{Blocking sets on the complement of hyperplane arrangements in PG(n,q)}

In this section we generalize some interesting results on Blocking Sets in
$PG(n,q)$ to the complement of hyperplane arrangements.

Let $\Cal{A}=\{H_1,\ldots, H_m\}$ be an \textit{arrangement of hyperplanes} 
in $PG(n,q)$ and
\begin{equation*}
M(\Cal{A})=PG(n,q) \setminus \cup_{i=1,\ldots m} H_i
\end{equation*}
be the \textit{complement} of the arrangement.

Let $\Cal{A}_a=\{H_1,\ldots,H_m\}$ be an arrangement in $AG(n,q)$ and
\begin{equation*}
M_a(\Cal{A})=AG(n,q) \setminus \cup_{i=1,\ldots m} H_i
\end{equation*}
be the complement in the affine space.

Given an arrangement $\Cal{A}$ in $PG(n,q)$ ($AG(n,q)$), the 
\textit{corresponding arrangement} in $PG(k,q)$ ($AG(k,q)$) for $k \neq n$
is the arrangement given by hyperplanes with the same equations of those in
$\Cal{A}$. From now on we will also use $\Cal{A}$ to refer to corresponding
arrangements of $\Cal{A}$. 

Let us remarks that the statements of propositions \ref{sottospazi}, 
\ref{sovraspazi} and 
corollary \ref{somme} above hold trivially also for Blocking sets in 
the complements $M(\Cal{A}) \subset PG(n,q)$ and 
$M_a(\Cal{A}) \subset AG(n,q)$.

Moreover we have the generalization of Mazzocca-Tallini's theorem:

\begin{thm}\label{MTal}Let $M(\Cal{A})$ be the complement of an 
arrangement $\Cal{A}$ in the projective (affine) space $PG(n,q)$ 
($AG(n,q)$). Then we can find a function $b_{p,\Cal{A}}(t,q)$
($b_{a,\Cal{A}}(t,q)$), depending on t and q, such that 
$M(\Cal{A}) \subset PG(n,q)$ ($M(\Cal{A}) \subset AG(n,q)$) 
has t-blocking sets if and only if $r \leq b_{p,\Cal{A}}(t,q)$ ($r \leq b_{a,\Cal{A}}(t,q)$).
\end{thm}

\textbf{Proof} Let $d_{M(\Cal{A}(n,q))}$ be the maximum of 
the dimensions of linear subspaces in $M(\Cal{A}(n,q))$.

Then, by simple geometric consideration, we remark that
$\{d_{M(\Cal{A}(n,q))}:n \in \N \}$ is an increasing function.

Let us suppose that $\{r_n:n \in \N \}$ is a sequence of integer such that
for all $r_n$ there is an $h$-blocking set $B(r_n)$ in $M(\Cal{A}(n,q))$.

Then for $d_{M(\Cal{A}(r_n,q))}>h$ let $B(r_n) \cap S(r_n)$ be the 
intersection between $B(r_n)$ and a linear variety $S(r_n)$ in 
$M(\Cal{A}(r_n,q))$ of dimension $d_{M(\Cal{A}(r_n,q))}$.
Then $B(r_n) \cap S(r_n)$ is an $h$-blocking set in $S(r_n)$, but this 
contradicts theorem \ref{funzione importante}. $\quad \square$\\
\\

\textbf{The braid arrangement}\\

Let us consider the \textit{Braid arrangement}
$$\Cal{A}(A_{n,q})=\{H_{i,j}\}_{1 \leq i < j \leq n+1}$$ given by reflection
hyperplanes $H_{i,j}$ in $PG(n+1,q)$ of equations 
$\alpha_{i,j}:x_i-x_j=0$.

The complement $M(\Cal{A}(A_{n,q}))$ is defined by points
$$(x_0,x_1,\ldots ,x_{n+1}) \in PG(n+1,q)$$ such that $x_i \neq x_j$,
$1 \leq i < j \leq n+1$. Then we have the following:

\begin{prop}\label{arratrecce}The complement $M(\Cal{A}(A_{n,q}))$ contains 
Blocking sets for all $n \leq q-1$ and it is empty otherwise.  
\end{prop}

In order to prove proposition \ref{arratrecce} we need the following lemmas:

\begin{lemma}\label{numerorette} For any two points 
$\underline{x}$ and $\underline{y}$  in the complement $M_a(\Cal{A}(A_{q,q}))$,
the line $$\underline{y}+t(\underline{x}-\underline{y})$$
is in the complement iff $\underline{x}-\underline{y}=(1,\ldots,1)$. 
In particular $M_a(\Cal{A}(A_{q,q}))$ contains $(q-1)!$ lines.
\end{lemma}

\textbf{Proof.}Clearly $M_a(\Cal{A}(A_{q,q}))$ contains all lines trough a 
point $P \in M_a(\Cal{A}(A_{q,q}))$ and with direction $(1,\ldots,1)$.

By contrary let $\underline{x}, \underline{y} \in M_a(\Cal{A}(A_{q,q}))$
be such that $\underline{x}-\underline{y} \neq (1,\ldots,1)$. 
If $\underline{x}=(x_1,\ldots,x_{q+1})$ and
$\underline{y}=(y_1,\ldots,y_{q+1})$ then there exists $i,j$ such that 
$x_i-y_i \neq x_j-y_j$. Let define:
\begin{equation*}
t_0=\frac{(y_j-y_i)}{(x_i-y_i)-(x_j-y_j)};
\end{equation*}
then $t_0 \neq 0$. Moreover, if $P_i$ and $P_j$ are, respectively, the 
$i$-th and the $j$-th entries of
$P=\underline{y}+t_0(\underline{x}-\underline{y})$, then $P_i=P_j=
\frac{x_iy_j-x_jy_i}{(x_i-y_i)-(x_j-y_j)}$, i.e. 
$P \notin M_a(\Cal{A}(A_{q,q}))$ $\quad \square$

\begin{lemma}\label{arraffine}$M_a(\Cal{A}(A_{q,q}))$ contains Blocking Sets.
\end{lemma}

\textbf{Proof.} Let us remark that $M_a(\Cal{A}(A_{q,q}))$ contains
$q!$ points and $(q-1)!$ lines with the same direction $(1, \ldots, 1)$ (by
lemma \ref{numerorette}). It follows, trivially, that the set given by one 
point for any line is a Blocking set. $\quad \square$\\
\\

\textbf{Proof proposition \ref{arratrecce}.} As consequence of 
lemma \ref{arraffine} and theorem \ref{MTal} the complement 
$M_a(\Cal{A}(A_{n,q}))$ contains Blocking sets if $n \leq q$.\\
Moreover $PG(1,q)$ contains blocking sets. Then the statement follows from 
corollary \ref{somme}.$\quad \square$

\begin{rem} Let $M(\Cal{A})$ be the complement of $\Cal{A}$ in $PG(n,q)$ 
($AG(n,q)$) and $V$ a subspace of $M(\Cal{A})$ without Blocking Sets. 
Then by proposition \ref{sottospazi} it follows that 
$M(\Cal{A})$ doesn't contain Blocking Sets.
\end{rem}

\begin{rem} This new theory on Blocking Sets gives rise to
a lot of questions and problems:
\begin{itemize}
\item how we can characterize a \textnormal{minimal unblocking arrangement} in
$PG(n,q)$ (or in $AG(n,q)$), i.e. an arrangement 
$\Cal{A}$ such that $PG(n,q)$ doesn't contain Blocking Sets while
$M(\Cal{A})$ does and such that $\Cal{A}$ is minimal with respect to the 
cardinality.\\
\item Vice versa, how we can characterize a \textnormal{minimal blocking 
arrangement}, i.e. an arrangement 
$\Cal{A}$ such that $PG(n,q)$ contains Blocking Sets while
$M(\Cal{A})$ doesn't and such that $\Cal{A}$ is minimal with respect to the 
cardinality.\\ 
An example of a minimal blocking arrangement is the line to infinity in
$PG(2,3)$. Indeed it is know that
$PG(2,3)$ contains Blocking Sets while $AG(2,3)$ doesn't.\\
\item Other interesting questions are:
\begin{enumerate} 
\item which is the link between the classical
Blocking Sets theory and the theory on the complement $M(\Cal{A})$ of an 
arrangement $\Cal{A}$?
\item Which is the link between the theory of arrangements on finite spaces and
the theory of blocking sets in the complement $M(\Cal{A})$ of an 
arrangement $\Cal{A}$?
\end{enumerate}
\end{itemize}
\end{rem}

\phantom{\cite{BoseBurton,82,BruenThas,Bruen1,173,OS,SzGaWe,Hirsc,Innam,Innamat}}
\newpage{\pagestyle{empty}\cleardoublepage}
\bibliographystyle{plain}
\addcontentsline{toc}{chapter}{Bibliografia}
\bibliography{articolo1} 

\end{document}